\documentclass[12pt,preprint]{aastex}

\newcommand{\appropto}{\,\rlap{\raise 0.5ex\hbox{$\propto$}}{\lower 
1.0ex\hbox{$\sim$}}\,}

\slugcomment{Ap. J. Lett., in press}

\shorttitle{M Dwarf HZ Planets Probably Lack Volatiles}
\shortauthors{Jack J. Lissauer}

\begin{document}


\title{Planets Formed in Habitable Zones of M Dwarf Stars Probably are Deficient in Volatiles}


\author{Jack J. Lissauer}
\affil{Space Science and Astrobiology Division, 245-3, NASA Ames Research Center, Moffett Field, CA 94035}



\begin{abstract}
Dynamical considerations, presented herein via analytic scalings and numerical experiments, imply that Earth-mass planets accreting in regions that become habitable zones of M dwarf stars form within several million years.   Temperatures in these regions during planetary accretion are higher than those encountered by the material that formed the Earth.  Collision velocities during and after the prime accretionary epoch are larger than for Earth.  These factors suggest that planets orbiting low mass main sequence stars are likely to be either too distant (and thus too cold) for carbon/water based life on their surfaces or have abundances of the required volatiles that are substantially less than on Earth.
\end{abstract}





\section{Introduction}

M (and early L) dwarfs are the lowest mass stars, and the smallest and least luminous members of the stellar main sequence.  They are also by far the most numerous class of stars in our galaxy, with $\sim$ 75\% 
of the stars in the extended solar neighborhood being M dwarfs \citep{reid2002}.  Thus, if planets orbiting M stars can be habitable, the number of potentially inhabited worlds within our galaxy could be much larger than if only planets orbiting solar type stars are capable of hosting life.  The presence of liquid water on a planet for long periods of time is considered to be a requirement for the origin and evolution of life as we know it.  It is shown herein that planets within M star habitable zones (HZs) formed rapidly while their star was quite luminous and are probably water-deficient. Thus, the dynamics of planetary accretion, together with the physics of stellar evolution, cast doubt on the suitability of planets orbiting M stars to host life.

Stars on the main sequence have relatively long periods of very slowly varying luminosity, potentially providing stable radiation fluxes for the origin and evolution of life.  Giant planets, brown dwarfs, and post-main sequence stars vary substantially in luminosity, and thus are far less conducive energy sources for life on the surface of a planetary body \citep[cf.] [for a possible exception involving a class of low-mass post-main sequence stars in the very distant future]{adams2005}. 

The main sequence luminosity of stars varies roughly as $M_\star^4$, where $M_\star$ is stellar mass. Thus, the locations of HZs (defined to be orbits in which liquid water can be present on the surface of an Earth-like planet) are much closer to low mass main sequence stars than they are to high mass stars \citep{kasting1993}.  As stellar nuclear fuel varies in proportion to $M_\star$, stellar lifetime goes as $M_\star^{-3}$, so stars more than about twice as massive as the Sun are probably not sufficiently long-lived to provide the energy for the development of advanced life. 

The lowest mass stars are very long-lived, but present a different set of challenges to life \citep{scalo2007}.  The most studied of these problems is that the proximity of HZs to very low mass stars implies that the rotation of a HZ Earth-like planet would be tidally altered to the synchronous state, with one hemisphere permanently illuminated and the other in the dark \citep{kasting1993}.  A thin atmosphere of such a planet would freeze out onto the planet's night hemisphere, but a sufficiently massive atmosphere could transport enough heat to the dark side to prevent atmospheric collapse, thus allowing habitable regions on the planet's lit hemisphere \citep{joshi1997}.  Perturbations from another planet could also maintain a HZ planet's orbital eccentricity against tidal locking, especially if the HZ planet lacked a substantial permanent deformation (and thus could rotate just slightly faster than synchronous, in equilibrium with the tidal forcing experienced by an eccentric planet).  

However, M dwarf stars present other challenges to planetary habitability.  As shown below, a particularly severe problem is the difficulty that a 1 M$_\oplus$ (Earth mass) planet in the HZ of an M dwarf has in accumulating and retaining water and other volatiles.  The Earth itself is volatile-poor.  Oceans and other near-surface reservoirs of H$_2$O comprise less than 0.03$\%$ 
of our planet's mass, with the mantle containing a comparable amount of water (K. Zahnle, personal communication, 2006).  Thus, Earth's total water abundance is very small compared to the cosmic rock:H$_2$O ratio of roughly unity. Earth accreted from planetesimals that condensed over a wide range of heliocentric distances \citep{wetherill1994}. The temperature distributions of protoplanetary disks are not tightly constrained.  Thus, the best estimates of the volatile contents of planetesimals as a function of heliocentric condensation distance are provided by analysis of meteorites that are analogous to asteroids of known heliocentric distance and by matching Earth's volatile inventory to planetary accumulation models \citep{morbidelli2000}.  Primitive meteorites that emanate from the inner asteroid belt (just beyond 2 AU) have water abundances $<  0.1\%$, whereas those from more distant regions of the asteroid belt can have more than 100 times this much water.  Accretion simulations imply that a nontrivial amount of material from the region of the main asteroid belt reached Earth, and that most of Earth's water probably came from planetesimals which condensed beyond $\sim$ 2.5 AU \citep{lunine2003, raymond2004}. Although current planetary accumulation models cannot be viewed as providing {\it ab initio} estimates of the accretion of water by terrestrial planets, models of this type are well-suited for comparing the relative amounts of volatiles accreted by terrestrial planets around stars of differing masses.

Stars are substantially more luminous during their formation epoch than they are once they reach the main sequence.  Low mass stars are significantly more luminous than their main sequence luminosities for longer than are solar mass stars \citep{stahler2005}, so ultimately habitable regions around M stars are hotter at young stellar ages than are regions that are to be subjected to the same stellar flux around sunlike stars. Additionally, energy diffusion is slower in inner protoplanetary disks because of the greater optical thickness, so there is likely to be an even larger excess of temperature within young circumstellar disks near smaller mass protostars. The location of the snow line within protoplanetary disks orbiting small stars \citep{kennedy2006} is thus more distant in proportion to the eventual location of the HZ than is the case for solar mass stars.

The timescales of accretion of planets of a given mass in the HZs of M stars are shorter than those of comparable planets around more massive stars.  This is because orbital periods are shorter, planetesimals are closer to one another and they occupy a larger fraction of their Hill spheres.  Moreover, orbital velocities are faster, implying higher impact speeds, so late accretion of volatile-rich bodies that condensed farther from the star may well remove more atmospheric gasses and water than they provide. Thus, while lower mass pre-main sequence stars remain more luminous than their main sequence luminosities for a longer time, Earth-mass planets forming in HZs around these stars accrete more rapidly, of material that is likely less water-rich, and within a dynamical environment in which they are more likely to lose atmospheric volatiles via impact erosion.  The dynamical characteristics of terrestrial planet accumulation around M stars are quantified in Section 2.  

\section{Accretion Timescales \& Impact Velocities}

For stars of solar mass and smaller, main sequence luminosity, $L_{\star}$, varies with stellar mass, $M_{\star}$, roughly as:
\begin{equation}
L_{\star} \approx \left ( \frac{M_{\star}}{{\rm M}_\odot} \right ) ^4{\rm L}_\odot \appropto  M_{\star}^4,
\end{equation}
where the symbol $\appropto$ signifies that the proportionality relationship is only approximate\footnote{The actual slope of the mass-luminosity relationship for 0.5 -- 1 M$_\odot$ stars is slightly steeper, between $M_{\star}^4$ and $M_{\star}^5$, but it flattens out to vary $\sim M_{\star}^2$ for lower masses \citep[and J. Scalo, personal communication, 2006]{hillenbrand2004}.  Therefore, the HZ is a little closer to K and early M stars ($M_{\star} > \frac{1}{3}$ M$_\odot$), than predicted using Eq. (1), but a somewhat farther from late M stars ($M_{\star} < \frac{1}{3}$ M$_\odot$) than given by this scaling.  }. Planetary temperature is proportional to $(L_{\star}/r_{\star}^2)^{1/4}$, where $r_{\star}$ is the distance from the star.

The distance of the HZ from the star, $r_{HZ}$, thus varies as:
\begin{equation}
r_{HZ} \propto L_{\star}^{1/2} \appropto M_{\star}^2.
\end{equation}
Terrestrial planet accretion proceeds until a stable configuration is reached \citep{lissauer1995, laskar2000}, and the separation of planets required for stability varies as $(M_p/M_{\star})^{2/7} r_{\star}$ \citep{wisdom1980}.  Thus, the surface mass density of the ensemble of solids within the protoplanetary disk, $\sigma$, required for local accretion of planets of a given mass (e.g., that of Earth) in the zone that becomes habitable varies as

\begin{equation}
\sigma \appropto \frac{M_{\star}^{2/7}}{r_{HZ}^2} \appropto r_{HZ}^{-13/7} \appropto M_{\star}^{-26/7}.
\end{equation}
When expressed in terms of surface density in the HZ, the value of $\sigma$ necessary for the formation of earthlike planets is, therefore, a very steep function of stellar mass.  But standard minimum mass models of our protoplanetary disk\footnote{These models are not constrained by data from regions as close to the Sun as the HZ distance is from an M star, because our Sun has no planets that orbit so near it. Also, flatter surface density models are derived in other minimum mass solar nebula models \citep{davis2005}, and such flat profiles better fit giant planet formation models \citep{lissauer1987} as well as most evolutionary models of protoplanetary disks.} \citep{weidenschilling1977} give $\sigma(r) \propto r^{-3/2}$, which is only slightly less steep than the $r_{HZ}^{-13/7}$ dependence in Eq. (3). Note that masses of protoplanetary disks around low mass stars are poorly constrained, but observations suggest that they typically are lower than those around 1 M$_{\odot}$ stars \citep{scholz2006, muzerolle2006}. If the surface densities of disks in ultimately HZs around small stars do not have substantially higher surface mass density than the minimum mass proto-solar disk at 1 AU, then planets in the HZs of M stars are small unless they accreted large amounts of material that migrated inwards from the outer disk or they themselves migrated inwards substantially.  Low-mass planets have shallower gravitational potential wells, and thus are more susceptible to atmospheric loss than are more massive planets. 

The growth rate of a terrestrial planet is proportional to the surface density of the disk and the orbital frequency; other factors related to gravitational enhancement in accretion cross-section also play a role \citep{safronov1969}, but this enhancement is of order unity during the final high-velocity phase of planetary growth, which dominates accretion timescales \citep{lissauer1987, agnor1999}.  Growth times thus scale as
\begin{equation}
T_{gr} \appropto \left(\frac{r_{\star}^2}{M_{\star}^{2/7}}\right)\left(\frac{r_{\star}^{3/2}}{M_{\star}^{1/2}}\right) = \frac{r_{\star}^{7/2}}{M_{\star}^{11/14}} \appropto M_{\star}^{6.2}.
\end{equation}
According to Eq. (4), Earth-mass planets that accrete within HZs of 0.5 M$_\odot$ stars should form in several million years; around 0.25 M$_\odot$ stars, the process should require $\lesssim$ 10$^5$ years.  Migration inwards requires the presence of a massive disk.   Optically thick disks around young stars have a broad range of lifetimes, although most disappear 1 -- 5 Myr after the stellar photosphere becomes visible (Figure 5 of \citet{briceno2007, meyer2007}).  Thus, whether it has grown {\it in situ} or migrated inwards from a greater distance, the planet should be in or very near its final orbit within the HZ within $10^7$ years of the star's formation.  

A slightly less steep scaling with stellar mass than given by Eq. (4)  is probably more appropriate to account for two physical effects: (1) The amount of gravitational focusing is reduced for accreting bodies which occupy a greater fraction of their Hill spheres \citep{greenzweig1990}.  (2)  The formulae used to derive Eq. (4) assume crossing orbits, whereas the very late stages of terrestrial planet growth are dominated by the time required for chaotic perturbations to excite sufficient eccentricities for continuation of orbital crossing.

The growth of terrestrial planets within the HZ of a $\frac{1}{3}$ M$_\odot$ star (which has a main sequence luminosity very close to (1/3)$^{-4}$ that of a 1 M$_\odot$ star \citep{hillenbrand2004})  was modeled numerically to test the scaling derived above.  The initial disk was analogous to that used for simulations of terrestrial planet growth around a 1 M$_\odot$ star as well as within binary star systems \citep[and references therein]{quintana2006}.  As in these previous studies, the simulations began with 140 `planetesimals' of mass 0.00933 M$_\oplus$ and 14 `planetary embryos' of mass 0.0933 M$_\oplus$, and the density of each body was taken to be 3 g/cm$^3$.  In order to leave the illumination of bodies that were at 1 AU in the initial disk unchanged according to Eq. (2) and maintain expected planetary sizes as per Eq. (3), the initial semimajor axes of these 154 bodies were scaled as 
\begin{equation}
a = \frac{a_p^{(3^{2/7})}}{9}, 
\end{equation}
where $a_p$ are the values used in the previous simulations referenced above and both $a$ and $a_p$ are in AU.  The other 5 initial orbital elements of each of the 154 accreting bodies were as in previous simulations.   Bodies were removed from the integrations if they approached closer to the star than 0.01 AU, or if their distance from the star exceeded 10 AU.  The integrations were performed using the hybrid symplectic integrator within the {\it Mercury} integration package \citep{chambers1999}.  The timestep was taken to be one hour. Deterministic chaos implies that integrations of this sort are valid in only in a statistical sense, so multiple numerical experiments were run. A pair of simulations, differing only by moving the initial position of one of the smaller bodies along its orbit by one meter, was performed without any giant planets, and four analogous runs were performed with giant planets similar to Jupiter and Saturn but with masses reduced by a factor of three and semimajor axes reduced by a factor of nine. 

Results of the four numerical simulations of planetary growth in the HZ of a $\frac{1}{3}$ M$_\odot$ star that has a pair of giant planets are displayed in Figure 1.  The final systems have similar numbers of terrestrial planets as those formed in simulations of planetary growth around 1 M$_\odot$ stars on which the initial disk conditions were based.  The final collision between two embryo-dominated bodies or loss of one such body occurred at 0.39 Myr, 0.38 Myr, 1.21 Myr and 0.67 Myr in these 2 Myr simulations; in comparison, the last reduction in number of embryos in the 31 simulations of terrestrial planet growth (each for a time span of 200 Myr) tabulated in \citet{quintana2006} occurred at times ranging from 18.7 Myr to 134 Myr, with a median final loss time of 78 Myr.  Growth rates around the $\frac{1}{3}$ M$_\odot$ star thus are more than 100 times as rapid as around the 1 M$_\odot$ star, which implies a somewhat less steep scaling than  Eq. (4), consistent with the above discussion.  The simulations lacking giant planets resulted in more final terrestrial planets extending to greater asterocentric distance, as in the case of a single star lacking in giant planet or stellar companions \citep{quintana2002}. In these runs, the final collision between two embryo-dominated bodies or loss of one such body occurred at 0.11 Myr and 0.18 Myr. Note that $\frac{1}{3}$ M$_\odot$ stars are more than ten times as luminous during their first $\sim 4$ Myr as they are when they reach the main sequence \citep{dantona1997}.

Planets grow by accretionary impacts.  Such impacts can remove previously accreted volatiles that reside in a planet's atmosphere.  Impact erosion is a favored mechanism for explaining the thinness of Mars's atmosphere relative to that of Earth \citep{melosh1989}.  Orbital speeds, and thus impact velocities, are higher in HZs around low-mass stars than they are at Earth, and such high collisional velocities may lead to more erosive impacts.  Earth is believed to have accumulated most of its water and other volatiles from solid bodies that condensed outwards of the orbit of Mars and were accreted by Earth towards the end of or subsequent to the major phases of our planet's growth \citep{morbidelli2000}.  Accretion from a similarly-scaled distance around other stars would result in approach velocities, $v_\infty$, that vary in proportion with the orbital velocity\footnote{The situation compares even less favorably for `local' impacts during the late stages of the primary growth of planets that reach the same masses, as they must be separated by a larger fraction of their orbital distance for stability about the smaller star.  Thus, an additional factor of $M_\star^{-2/7}$ must be included to account for the greater eccentricities of planetesimals which are cleared from the entire region between two growing planets, yielding $v_\infty \appropto M_\star^{-11/14}$.},

\begin{equation}
v_\infty \propto \left(\frac{M_{\star}}{r_{HZ}}\right)^{1/2} \appropto M_{\star}^{-1/2}.
\end{equation}
The larger ratio of the distance to the ice line to $r_{HZ}$ for M stars implies that fewer ice-rich planetesimals reach the HZ, and those that do have an even higher characteristic velocity than given by Eq. (6). The specific energy of an impact varies as the square of impact velocity. According to \citet{melosh1989}, substantial impact erosion of atmospheres requires collision velocities faster than twice the planet's escape speed.  Parabolic comets with prograde orbits in the ecliptic plane impact Earth at $< 2 v_e$.  However, all planetesimals with large semimajor axes impact a 1 M$_\oplus$ planet orbiting at $\frac{1}{9}$ AU from a $\frac{1}{3}$ M$_\odot$ star at $> 2 v_e$. Therefore, impacts in the HZs around small stars can significantly erode the atmospheres of $\lesssim$ 1 M$_\oplus$ planets. 

\section{Discussion}

In sum, under nominal circumstances, planets in main sequence habitable zones around M stars are likely to be fully formed and in their final orbits by the time the gaseous circumstellar disk has dissipated or several million years after planetesimal formation, whichever is later.  If growth is {\it in situ}, dynamical and thermal factors imply that the planets are unlikely to have large volatile inventories, and planetary masses are likely to be small.  The large collision speeds of impacting comets, as well as the high activity and luminosities of young M stars, may lead to substantial mass loss from planetary atmospheres, depleting any reservoirs of volatiles that planets within the HZs are able to accrete.

So are M dwarf stars totally unsuitable hosts for the development of advanced life?  No, there are various ways out of the difficulties pointed out above.  Theoretical estimates of planetary volatile inventories obtained from planet formation studies are not from first principles, but rather rely on normalization to the planets within our Solar System; as accretion models suggest large stochastic variations in volatile delivery \citep{obrien2006}, it is possible that the terrestrial planets within our Solar System are near the volatile-poor end of this distribution, and thus the normalization used herein is incorrect. A water-rich planet formed farther from the star \citep{kennedy2006} could migrate inwards to the HZ while the gaseous disk is still present; such a planet could initially have enough water that oceans could be retained even if it suffered significant losses during the star's young active phase.  A planet could be placed on an eccentric orbit that circularizes on the timescale it takes the star to reach the main sequence or somewhat longer, or be scattered inwards by a fellow planet after the star reaches main sequence and then be circularized by stellar tides.  But such scenarios require precisely the right amount of initial water or just the right dynamics.  Thus, while it is likely that some of the hundreds of billions of M dwarf stars in our galaxy have planets with temperatures, masses and compositions similar to Earth, the number of such planets is probably small, and Sun-like stars, despite being considerably less numerous, may well be the hosts of far more habitable planets.

\acknowledgments

I am grateful to Gennaro D'Angelo, Todd Henry, Dave Hollenbach, Jim Kasting, Mark Marley and especially to Gibor Basri for illuminating discussions and to John Scalo for comments on the manuscript.  Elisa Quintana provided assistance with the numerical calculations displayed in Figure 1.  This research was supported by NASA PG\&G, WBS 811073.02.01.01.12.

\clearpage

\begin{figure}
\plotone{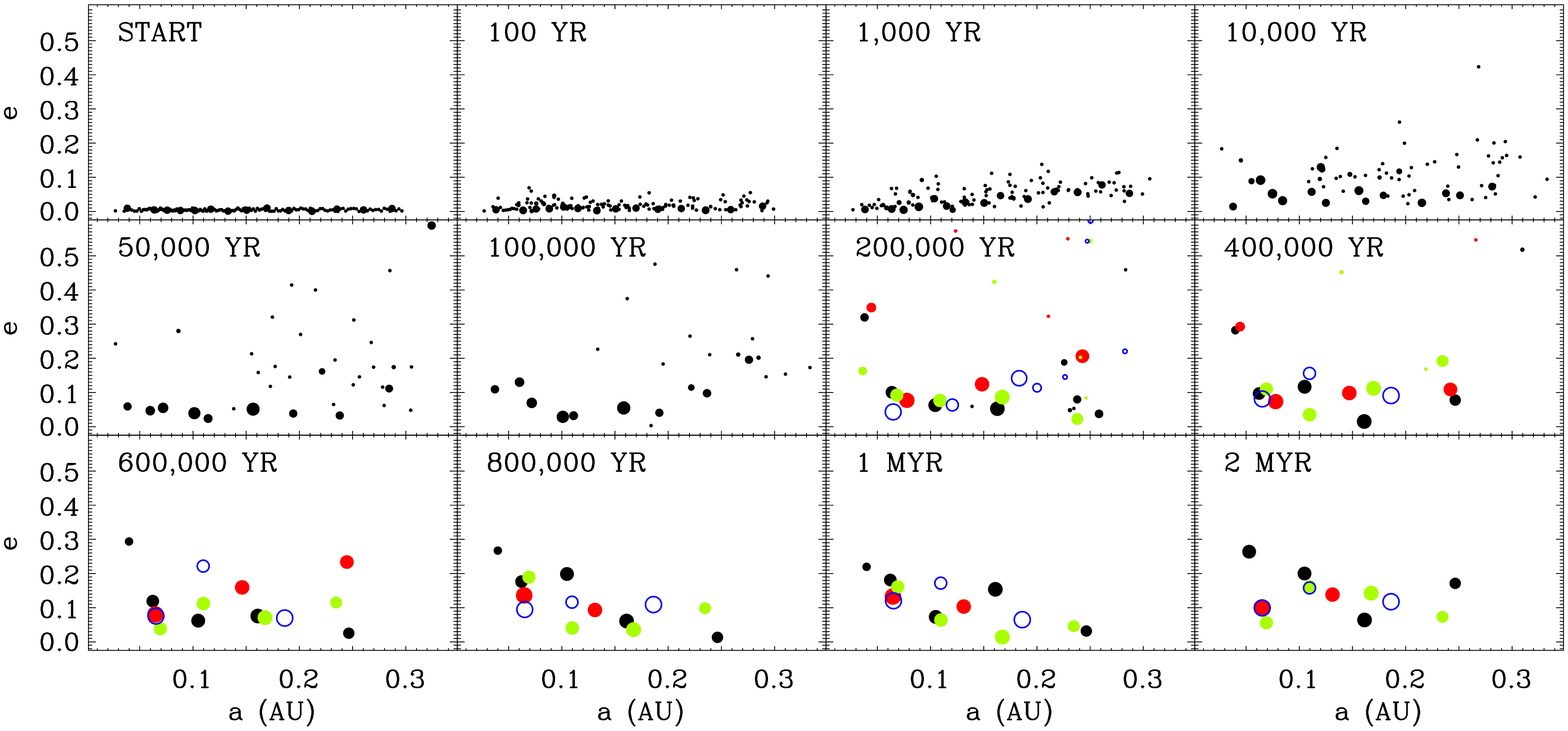}
\caption{The temporal evolution of four protoplanetary disks incorporating the ultimately habitable zone around a $\frac{1}{3}$ M$_\odot$ star.  Two giant planets, one-third the masses of Jupiter and Saturn, respectively, are also included.
  The planetary embryos and planetesimals are represented by circles
  whose sizes are proportional to the physical sizes of the bodies.  The evolution of one disk is shown in black at 12 different times. The green and red filled circles and blue open circles show the results at late times of the three runs with one planetesimal initially moved by 1 -- 3 m along its orbit.
The horizontal locations of the circles show the orbital semimajor axes of the bodies in the disk in AU, and vertical positions plot their
  eccentricities.  The initially dynamically cold disk heats up
  during the first 5 $\times$ 10$^4$ yr, especially in the outer region, where the
  perturbations of the giant planets
  are the greatest.  By  4 $\times$ 10$^5$ yr into the simulation, 3 -- 5 terrestrial planets have formed, with 0 -- 2 planetesimals remaining.  After 2 $\times$ 10$^6$ yr, each system contains 2 -- 4 terrestrial planets.  Note that the HZ is around 0.11 AU and that the largest planets formed have masses comparable to that of the Earth. \label{fig1}}
\end{figure}

\end{document}